# Providing content based billing architecture over Next Generation Network

**Mr. Kamaljit I. Lakhtaria**
*Lecturer, Atmiya Institute of Technology & Science, Rajkot*
*Email: kamaljit.ilakhtaria@gmail.com, (M): 9879515909*

**Abstract**
Mobile Communication marketplace has stressed that "content is king" ever since the initial footsteps for Next Generation Networks like 3G, 3GPP, IP Multimedia subsystem (IMS) services. However, many carriers and content providers have struggled to drive revenue for content services, primarily due to current limitations of certain types of desirable content offerings, simplistic billing models, and the inability to support flexible pricing, charging and settlement. Unlike wire line carriers, wireless carriers have a limit to the volume of traffic they can carry, bounded by the finite wireless spectrum. Event based services like calling, conferencing etc., only perceive charge per event, while the Content based charging system attracts Mobile Network Operators (MNOs) to maximize service delivery to customer and achieve best ARPU. With the Next Generation Networks, the number of data related services that can be offered, is increased significantly. The wireless carrier will be able to move from offering wireless telecommunications services to offering wireless telecommunication services plus a number of personalized Value Added Services like news, games, video broadcasts, or multimedia messaging service (MMS) through the network. The next generation Content Based Billing systems allow the operators to maximize their revenues from such services. These systems will enable operators to offer and bill for application-based and content-based services, rather than for just bytes of data. Therefore, the wireless business focus is no longer on infrastructure build-outs but on customer retention and increased average revenue per customer (ARPU). The mobile operator generates new revenues, strengthens brand value, and differentiates its service to attract and retain customers.

## 1. Introduction to Mobile Billing System

Mobile Network Operators (MNOs) have traditionally billed their customers on a strict usage basis, charging for service on a per-minute basis. But as wireless penetration slows and ARPU begins to lag, carriers are looking for ways to recoup the cost of their network upgrades to 2.5G and 3G [1]. Wireless data and content-based services are being touted as a way for service providers to differentiate themselves and reduce customer churn, but in order to make any money, carriers need to upgrade their billing systems to support content-based services. Mediation technology collects usage information from the network and presents it to the billing system, allowing carriers more flexible billing options, such as billing for Web sites, emails, applications such as streaming media, etc.

### 1.1 Billing types

Mobile billing in general terms states the system in which calculation and collection of monetary compensation is taken from the customers for using the telecommunication services [2]. Customer base for Mobile Billing Payment can be categorized in two broad categories like Post Paid and Pre Paid. But the ultimate concept of Billing is a part from this, because billing of Mobile Communication usage depends on type of use, and as per use Mobile Services broadly divide in two categories:

- Event based Billing
- Content based Billing

## 1.2 Event Based Billing

Mobile User when using Mobile for voice call, SMS sending for each event one particular charge is added into his account, For Next Generation Network a call detail record (CDR) is maintained by the service provider of each and every customer. All Events are recorded with a unit like voice call as per local call and roaming charge and all such so that as per usage of total unit a fix charge is multiplied [3].

Event based billing can no longer meet the needs of the new data services that have been launched lately. Content-based billing introduces different billing rate tables based on the value of the data content to be delivered. In this case, the operator can charge for both basic network access services and value-added services.

## 2. Content based services: at a Glance

The modern content based billing provides customers with various benefits out of which the biggest advantages of the content based billing is that mobile operators allow customers to pay or prepay for different types of content and services. The services of content-based billing include Websites, individual files, Multimedia Message Service (MMS), and lots more. In this fast age of technology, content-based billing has great significance and it is considered as a very profitable business. The mobile companies charge different standard of fees for the different types of content and services that are offered to customers [4].

**Table 1: Content and billing method billing**

| Content/Service Provided | Billing Method |
|---|---|
| • User self-service<br>• Balance inquiries<br>• Account replenishment<br>• Service activation | Free |
| • Local hosted multimedia<br>• MP3 files | Per byte |
| • News<br>• Stock quotes<br>• Sports scores<br>• Weather | Per click |
| • Personalization<br>• Ring tones<br>• Screen savers | Per download |
| • Internet games | Per game |
| • Event broadcasts<br>• Movies<br>• Interviews<br>• Sports videos | Per event, with separate quote request for each event |
| • E-mail<br>• Post Office Protocol (POP)<br>• Internet Message Access Protocol (IMAP)<br>• Simple Message Transport Protocol (SMTP) | Per byte |
| • Streaming media<br>• Real-Time Streaming Protocol (RTSP) | Per second |

## 3. Requirements On Charging Schemes

Charging for today's VASs consists of a charge for the basic service and a premium for the added value. Both charges are based on the duration of the call. In the future, due to the greater variety of services on offer, more flexible charging schemes for the premium would be desirable. Flexibility relates to the parameters, which determine the charge (in addition to the duration of the call, the charge could depend on the amount of data transferred), to the variety of different possible tariffs, and to the ease with which a certain tariff can be changed.

The value of a particular piece of information retrieved by a user from a VASP at any one time may be quite small. Charging schemes should thus not require a large financial overhead in order to process the charge. Furthermore, the scheme should have a performance compatible with the requirements of a mobile system. In short, the charging scheme must be efficient.

It is expected that the evolution of current mobile systems towards NGN will also see the emergence of many new network operators, NGN service providers, and VASPs, which may have serious implications for the trust relationships between them.

Thus, the charging scheme must be secure against cheating (such as overcharging by the VASP or underpaying by the user), and the parties involved should have the assurance that justified claims relating to charges can be proved and that unjustified claims cannot be successfully made. This is often called incontestable charging.

## 4. Content Based Charging for Next Generation Networks

Wireless service providers in today's world realize the importance of always-on wireless packet data networks enabling many new service offerings. Mobile users are willing to pay a premium for the content and services they are interested in. those services are video broadcast, multimedia services, news, gaming, internet browsing, music download and many more [5].

Content Based Billing (CBB) not only allows service providers to bill for application-based (IP) and content-based services, but also allows them to generate additional revenue by differentiating their service charges. The service providers, with CBB functionality, will be able to charge a premium for selected data transfers, monitor the development and usage of mobile content and differentiate their mobile data charges to gain a competitive advantage.

Hence, a Content Based billing (CBB) system is a vital element of wireless data business development.

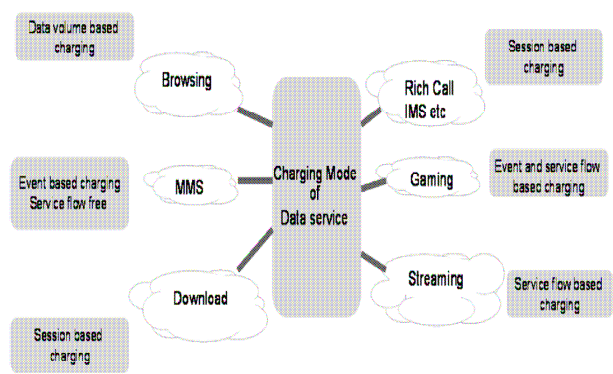

*Figure 1: Data service charging modes in Mobile communication Network*

### 4.1 Content Based Billing (CBB)

NGN networks like GPRS and UMTS networks; the GGSN (Gateway Support node) plays a significant role in the rollout of next generation networks. It provides a primary interface between the carrier's cellular network and the IP services layer. The GGSN is capable of offering enhanced service creation, billing, and IP traffic management by using its full visibility of the sessions' bearer traffic. These functionalities enable the collection of data and the billing of newly launched services, making it possible for the operator to measure the rate of adoption. The Content Based Billing (CBB) feature offered by C-GGSN enables differentiation between the various data flows, allowing different billing models to be applied. Content Based Billing solution provides key functionality including support of per-subscriber personalized IP packet filtering and IP flow based recording and reporting.

Content Based billing uses the same technology as the GGSN firewall functions, to filter traffic into flows based on IP quintuples (source IP address, destination IP address, port numbers, protocol and/or url). The operator can configure a set of patterns where each pattern corresponds to a rating bucket. When traffic passing through the GGSN matches a rule, the data packet will be accumulated in the related rating bucket. The CBB feature supports both post-paid and pre-paid users. For prepaid users, the GGSN demands coupons in real-time to charge for the traffic and gates the corresponding IP flow. Coupon acquisition and return is passed to the online charging systems for rating, credit checking and control. The CBB feature supports both post-paid and pre-paid users. For prepaid users, the GGSN demands coupons in real-time to charge for the traffic and gates the corresponding IP flow. Coupon acquisition and return is passed to the online charging systems for rating, credit checking and control.

### 4.2 Architecture for CBB over Next Generation Networks

The operator can define a special service-charging rate (from zero charging rate through full charging rate) for a specific content-based billing based on service market policy, price mechanism, and relationship with the Internet content provider. The Call Detail records (CDRS) generated by CBB system can be delivered through G-CDR extension attributes over the Ga interface, or be used directly to generate billing records according to charging rate information set by the user on the GGSN.

The GGSN can report billing records or information over different interfaces: [6]
- GTP (Ga interface) as in GTP accounting to Charging Gateway function (CGF),
- IETF RADIUS messaging (Gi interface) as in RADIUS accounting to AAA servers
- TCP/IP based Diameter protocol (Gy interface) as in real-time Prepaid. Together with a Prepaid Server, which could be an SCP and/or an online Charging System, the GGSN offers an integrated real-time pre paid solution that facilitates roaming.

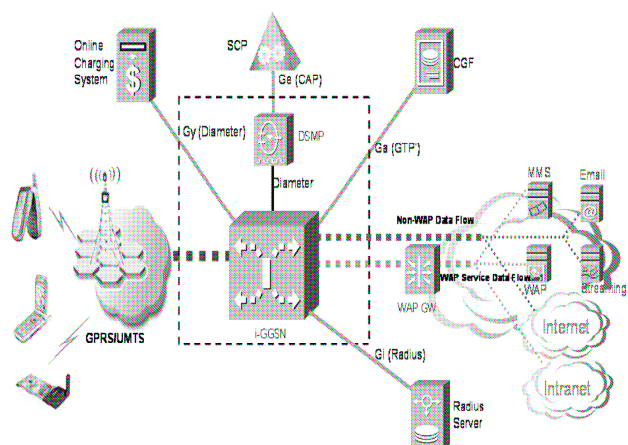

*Figure 2: Content Based Service delivery architecture over NGN*

## 4.3 Components of NGN architecture for CBB management

For accurately managing various Content based services Next Generation Networks enables two components for securing and accuracy for CBB. Two main components: the C-GGSN and the DSMP.

### 4.3.1 C-GGSN functions
- This component works to monitoring Deep packet inspection DPI, transfers between Value added services provide to User as per request send by users.
- GCSN also works to manage online/offline-charging architecture and service management.
- The heart of all service, Service authentication and Service control with Service based QoS control providing also enabled by with C-GGSN component.

### 4.3.2. DSMP functions
- DSMP works to manage common charging for customer as well as service providers as CBB provides various service in various charging category
- The Service directory center also enhance appropriate service request from customer sending to appropriate Value added service provider in NGN Network, along with service authentication task also performed in DSMP which integrates Common provisioning of Service providers.

## 4.4 Billing Recording and Generation

C-GGSN enables rich billing functions; it generates the GGSN Call Detail record (G-CDR). The GCDR is a data service billing record generated by the C-GGSN, which records the billing information relating to the external network usage. These records will be sent via Ga interface to the charging gateway for handling. The processed Call Detail records (CDRs) are then forwarded to the billing center for further processing.

Billing begins after a mobile user activates a PDP context, which creates and opens a billing record. When the PDP context is deactivated, the billing record is closed and billing stops. Each activated PDP context has a corresponding G-CDR billing record.

Billing methods can be classified as follows:
- GTP accounting and hot billing: G-CDRS
- RADIUS accounting: RADIUS accounting messages
- GGSN integrated Prepaid: Coupons (for time or data volume)

A G-CDR can have several containers. The GGSN generates new accounting containers under the following conditions:
- Volume counter limit is reached for the specific PDP session and the volume counter limit is configured through SPU in the APN profile.
- QoS profile changes when the GTP version remains unchanged.
- Configurable times of the day based on a TOD Profile. An APN profile may be assigned a TOD Profile allowing for the generation of a new container at any hour of the day.

## 4.5 CBB Services and Filtering Rules

When Content-Based Billing (CBB) is applied to a user session, all packets flowing between the user terminals and the destination content/application servers are subject to inspection and categorization into different flows.

A flow is made up of packets that match a particular packet-filtering rule. An IP packet flow is typically characterized by the IP quintuple flow identification:

- Layer source and destination IP addresses
- Layer 4-7 protocol types
- Source and destination port (TCP/UDP) numbers the GGSN uses the IP 5-tuple-flow identification as the basis for filtering IP flows.
- However, there is IP packet flows that require more than this 5-tuple filtering, as they involve application protocols that exert complex or static behaviors, and involve complex addresses and multiple port channels.
- These flows call for deep packet inspection. The C-GGSN is capable of filtering beyond the basic 5-tuple filtering. The general IP stack with its overlaying application protocols is shown in figure below.

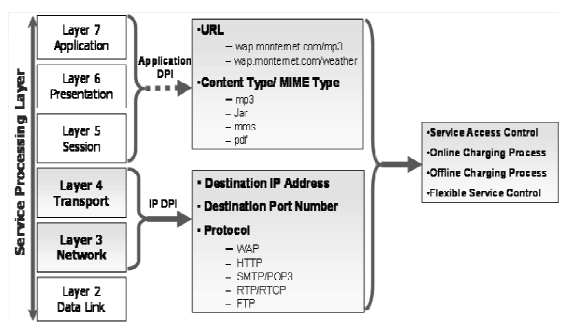

*Figure 3: NGN enabled C-GGSN Filtering Function*

## 5. Secured Charging Model for CBB over NGN

The relationship between the entities in NGN enabled network charging scheme is illustrated in Figure 5.

The VASP (Value Added Service provides) information to the user and sends corresponding charge requests. As the information is being provided, the user pays for the service by sending payment tokens. The VASP is able to check the validity of these tokens based on a certificate of the user's credentials issued by the NGN service provider, or an appropriate TTP acting on behalf of the NGN service provider.

As part of the offline payment clearance procedure, the VASP forwards billing information proving the claims on the user to the user's NGN service provider, who in turn bills the user as part of the regular subscriber billing process. In return the user sends a payment to the NGN service provider, who forwards the due share to the VASP.

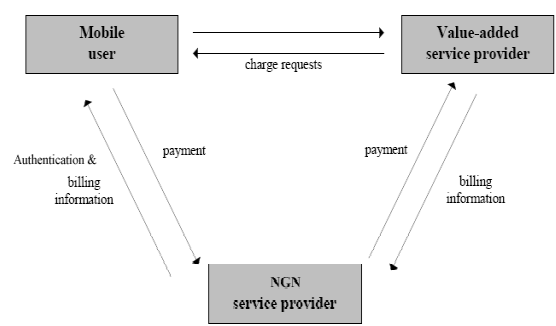

*Figure 5: Secured Charging Model*

## 6. Benefit to mobile operator

With the increasing demand from customer, Mobile Network operators must adapt Content service system, the content based billing system work not only to provide billing architecture as well as support to trap customer's choice of content with this in future new service of customer's choice also served to customers.

For Mobile Operators advantages with Content based billing are:

- Increase revenue potential through more granular billing
- Realize revenue sooner by offering prepaid services
- Differentiate the brand to strengthen customer loyalty
- Move up the value chain by offering services in addition to simple access

## 7. Conclusion

Mobile Communication system not limited to Voice and text based communication; customer needs to share their feeling, expressions over mobile equipment. Mobile enables entertainment anytime, anywhere to anyone of his or her users. But this happens only with sharing of Content of various categories over Next Generation Mobile Network. But for managing various types of content and specially adapt separate method of charging for each service, the Content Based Billing system helps operators will be able to develop billing structures to maximize the revenue and their competitive position in the fast growing and dynamic market. By automatically gathering data on content and usage, the CBB solution helps the operator to develop and manage content partnerships quickly and effectively in an environment where the end users place more demands on innovative and useful mobile data services. This agility is a key success factor in the Mobile Data business. That's no doubt with Mobile usage customer always aims best experience of their choice of content, but serving customer with best QoS and QoE the Content based billing system enhance will be compulsory for Mobile operator to maximize their revenue by deploying such a feature into their network